\documentclass[10pt,prb,aps,twocolumn]{revtex4}

\usepackage[dvips]{graphicx}

\begin{document}
\title{Noise-induced effects in high-speed reversal of magnetic dipole}

\author{A.L. Pankratov, S.N. Vdovichev, I.M. Nefedov}
\email[]{alp@ipm.sci-nnov.ru}
\affiliation{Institute for Physics of
Microstructures of RAS, Nizhny Novgorod, 603950, Russia}

\begin{abstract}
The effect of noise on the reversal of a magnetic dipole is
investigated on the basis of computer simulation of the
Landau-Lifshits equation. It is demonstrated that at the reversal by
the pulse with sinusoidal shape, there exists the optimal duration,
which minimizes the mean reversal time (MRT) and the standard
deviation (jitter). Both the MRT and the jitter significantly depend
on the angle between the reversal magnetic field and the anisotropy
axis. At the optimal angle the MRT can be decreased by 7 times for
damping $\alpha$=1 and up to 2 orders of magnitude for
$\alpha$=0.01, and the jitter can be decreased from 1 to 3 orders of
magnitude in comparison with the uniaxial symmetry case.
\end{abstract}

\maketitle

The decrease of sizes of magnetic nanoparticles used in storage
media leads to the increase of fluctuations and, therefore, to
increase of storage and switching errors (jitter). Thus, theoretical
investigation of noise-assisted high-speed switching of magnetic
dipoles is of crucial importance. The most studies are based on
computer simulation of the Landau-Lifshits equation \cite{AA}
describing the dynamics of the magnetic dipole in a magnetic field.
It has been found that there is an optimal angle between the applied
magnetic field and the anisotropy axis, which is typically around 45
degrees \cite{he},\cite{lyb02}; with the decrease of the magnetic
field rise time, the coercivity of magnetic particle (dynamic
coercivity) increases \cite{wel}, \cite{bert0}. The influence of the
damping and the external magnetic field amplitude on the reversal
time has been investigated in \cite{he},\cite{bert1},\cite{bert3};
the dependence on the pulse shape of the field has been studied in
\cite{switch}. All the above mentioned results have been obtained at
zero temperature, i.e. without account of thermal fluctuations.
However, at finite temperature the most important problem is the
stable magnetic reversal: the remagnetization process must occur
with minimal switching time and the standard deviation.
Unfortunately, there were very little number of works, devoted to
investigation of statistical characteristics of magnetic reversal
processes. The mean reversal time (MRT) has been studied in Ref.s
\cite{mst}, \cite{mst2}, and in Ref. \cite{mst2} it has been shown
that during noise-assisted reversal, the noise leads to the decrease
of MRT.

In the present paper the investigation of the reversal process of a
magnetic dipole has been performed on the basis of computer
simulation of the Landau-Lifshits equation with thermal fluctuations
taken into account. It is focused on the investigation of
statistical characteristics of the reversal process with the aim to
find an optimal regime of reversal with the smallest mean reversal
time and the standard deviation.

The dynamics of magnetic dipole is described by the Landau-Lifshits
equation:
\begin{equation}
\frac{d\overrightarrow{M}}{dt}=-\frac{\gamma}{\beta}\left[\overrightarrow{M}\times
\overrightarrow{H}\right]-\frac{\alpha\gamma}{\beta M_s}
\left[\overrightarrow{M}\times\left[\overrightarrow{M}\times\
\overrightarrow{H}\right]\right], \label{LL}
\end{equation}
where $\overrightarrow{M}$ is the magnetization of a particle,
$\overrightarrow{H}$ is the effective magnetic field, $\gamma$ is
the gyromagnetic constant, $\beta=1+\alpha^2$, $\alpha$ is the
damping, $M_s=\left|\overrightarrow{M}\right|$ is the saturation
magnetization. The effective magnetic field contains the following
components:
$\overrightarrow{H}=\overrightarrow{H_a}+\overrightarrow{H_e}+\overrightarrow{H_T}$,
where $\overrightarrow{H_a}$ is the anisotropy field,
$\overrightarrow{H_e}$ is the external field, and
$\overrightarrow{H_T}$ - fluctuational field. The fluctuational
field is assumed to be white Gaussian noise with zero mean and the
correlation function:
$\left<H(t)_{Ti}H(t')_{Tj}\right>=\frac{2\alpha kT}{\gamma M_s
V}\delta(t-t')\delta_{ij}$, where $k$ -- Boltzmann constant, $T$ is
the temperature, and $V$ is the volume of the magnetic particle.

Let us consider the reversal of magnetic dipole, initially
magnetized along anisotropy axis and along $x$-axis from the state
$\overrightarrow{M}[+M_s,0,0]$ to the state
$\overrightarrow{M}[-M_s,0,0]$. To find the area of parameters where
the fastest and the most reliable reversal occurs, as the
characteristic to be studied let us choose the first passage time of
a certain boundary. The mean first passage time (the mean reversal
time, MRT) $\tau$, and the standard deviation of the first passage
time $\sigma$ (SD, jitter) are \cite{ACP}: $\tau$=$\langle t
\rangle$=$\sum_{i=1}^{N}t_i/N$, $\langle t^2
\rangle$=$\sum_{i=1}^{N}t_i^2/N$, $\sigma$=$\sqrt{\langle t^2
\rangle-\langle t \rangle^2}$, where $t_i$ is the first passage time
of an absorbing boundary and $N\ge 10000$ is the number of
realizations. As in Ref. \cite{bert1} let us choose the boundary as
the passage of the point $\overrightarrow{M}[0,M_y,M_z]$.

In the calculations it is convenient to use the parameters, related
to the magnetic recording media \cite{lyb02}: $\alpha$=$0.1$,
$\gamma$=$1.76\cdot 10^{7}{\rm Hz/Oe}$, $M_s$=360 ${\rm Oe}$,
$V$=$2\cdot10^3{\rm nm^3}$, the anisotropy constant $K$=$7.2\cdot
10^5 {\rm erg/cm^3}$. The static coercivity is $H_c$=$2K/M_s$=4000
${\rm Oe}$. For modeling we take the amplitude of the magnetic field
to be $H_0$=6000 ${\rm Oe}$. It is known that the driving by the
signal with sharp fronts leads to the minimal MST \cite{PRL}.
However, the pulses used in real recording media systems have finite
rise time \cite{lyb02}. As an example of a driving with smooth
fronts we consider the sinusoidal pulse
$\vec{H_e}$=$\vec{e}H_0\sin\pi t/t_p$ with the width $t_p$ (see the
inset of Fig. 1), where $\vec{e}$ is unitary vector of magnetic
field direction. If the switching during $t_p$ does not happen, the
computation is continued for $\overrightarrow{H_e}$=0 until some
long period of time $t_f$, much larger than any other relaxation
time scale. Our aim is to find the parameter range, for which the
reversal by a smooth pulse takes place quickly and reliably.

The Landau-Lifshits equation with noise has been computed both by
the Heun method programmed in Fortran and by the specialized package
SIMMAG (Simulation of MicroMAGnets), developed in the laboratory of
mathematical modeling of IPM RAS.
\begin{figure}[h]
\resizebox{1\columnwidth}{!}{
\includegraphics{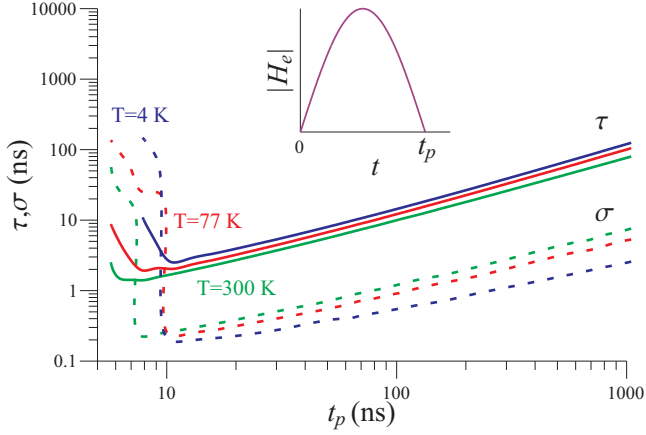}}
\caption{The mean reversal time (solid curves) and the standard
deviation (dashed curves) versus pulse width for the zero angle
between anisotropy axis and the driving field $\theta$=0. Inset: the
driving pulse. } \label{fig1}
\end{figure}
\begin{figure}[h]
\resizebox{1\columnwidth}{!}{
\includegraphics{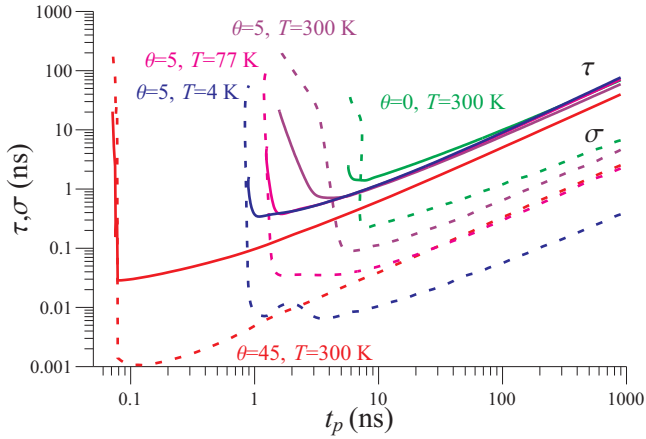}}
{\caption{The MRT (solid curves) and SD (dashed curves) versus pulse
width for different angles between anisotropy axis and external
field and different temperatures for $\theta=5^\circ$. }
\label{fig2}}
\end{figure}
\begin{figure}[h]
\resizebox{1\columnwidth}{!}{
\includegraphics{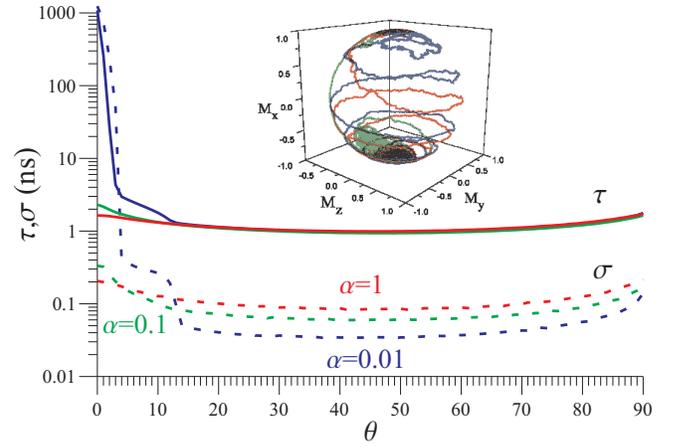}}
{\caption{The MRT (solid curves) and SD (dashed curves) versus the
angle between anisotropy axis and external magnetic field for
different values of damping $\alpha$, $t_p=8{\rm ns}$ and
temperature $T$=300 K. Inset: trajectories of the magnetization for
different angles, $\alpha$=0.1, $t_p$=15.7 ns, $T$=300 K;
$\theta=0^\circ$ - blue curve, $\theta=5^\circ$ - red curve,
$\theta=45^\circ$ - green curve.} \label{fig3}}
\end{figure}

It is known that for zero temperature $T=0$ the reversal of the
dipole by the longitudinal field, $\theta=0$, does not occur, since
the dipole is in the equilibrium state, even if this state is
unstable. The presence of thermal fluctuations allows to move the
dipole away from this equilibrium state. In Fig. 1 the plots of the
mean reversal time $\tau$ and the standard deviation $\sigma$ are
presented. First of all, it is seen that both $\tau$ and $\sigma$
have minima as functions of the driving pulse width. This indicates
that both these temporal characteristics can be minimized by the
optimal choice of pulse duration. Similar effect has recently been
observed for Josephson junctions \cite{PRL}. The decrease of the MRT
at large durations is due to the fact that with decrease of the
width the potential barrier disappears faster. With further
shortening of the pulse, the magnetization does not have enough time
for the complete reversal during $t_p$, so the MRT increases. This,
actually, means that for rather short pulses the transition occurs
due to effect of fluctuations (the so-called noise-induced
switching). The standard deviation with decrease of the pulse width
behaves similarly to $\tau$, but the curves for different
temperatures cross each other. For long pulses smaller temperature
leads to smaller $\sigma$, but for short pulses - vice versa. This
is also explained by the transition from the regime of switching by
the external field to the noise-induced escapes, since it is
well-known, see, e.g., \cite{ACP}, that in the noise-induced regime
the SD is approximately equal to the MRT.
\begin{figure}[h]
\resizebox{1\columnwidth}{!}{
\includegraphics{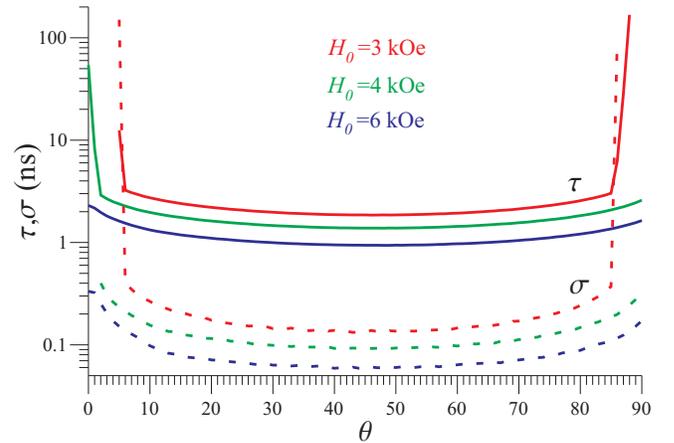}}
{\caption{The MRT (solid curves) and SD (dashed curves) versus the
angle between anisotropy axis and external magnetic field for
different values of magnetic field amplitude, $t_p$=8 ns and
temperature $T$=300 K.} \label{fig4}}
\end{figure}

Let us point out few more peculiarities, which are clearly visible
in Fig. 1: higher temperature leads to smaller MRT, i.e. noise
allows to speed up the reversal, which agrees with the predictions
of Ref. \cite{mst2}, but contradicts with the results of Ref.
\cite{PRL}. Besides, the dependence $\sigma\sim\sqrt{T}$ usual for
Josephson junctions at large pulse widths, is not also reproduced
here, from Fig. 1 one can see that $\sigma(T)$ dependence is slower
than $\sqrt{T}$. The explanation of such an unusual behavior is the
specific nature of the reversal process at $\theta=0$: in difference
with the Josephson junctions \cite{PRL}, due to location of initial
condition at the unstable equilibrium point, the reversal is
impossible at zero temperature and the deterministic reversal
trajectory does not exist. That is why fluctuations help to leave
the unstable initial state and namely this leads to the described
above peculiarities. It should be noted that the above investigated
case $\theta=0$, which is mostly studied analytically
\cite{kalm95}-\cite{kalm97}, is degenerate: first, technically the
angle between the anisotropy axis and the external field can be set
up with a certain precision; second, the MRT is largest in this
case, see below.

In Fig. 2 the MRT and SD are given for three different values of
angles $\theta$=$0^\circ,5^\circ,45^\circ$ for $T$=300 K and three
different temperatures for $\theta$=$5^\circ$. First, let us focus
on the curves for $\theta$=$5^\circ$. In spite that at large $t_p$
one can see little decrease of MRT with increase of the temperature,
at small $t_p$ around minimum the opposite effect of noise delayed
switching is clearly visible, quite similar to the one observed
before for Josephson junctions \cite{PRL}, and in general case of
nonlinear systems \cite{agud}. Besides, here the SD behaves as
$\sigma\sim\sqrt{T}$, see Ref. \cite{PRL}.

From Fig. 2 it is obvious that for $\theta$=$45^\circ$ the reversal
is faster and more stable than for $\theta$=$0^\circ$,$5^\circ$ at
all other equal conditions. Besides, the difference between MRT for
the angles $\theta$=$0^\circ$ and $\theta$=$5^\circ$ is two times,
while the difference of SD is about three times. For the cases
$\theta$=$5^\circ$ and $\theta$=$45^\circ$ the gain is even larger,
more than one order for MRT and almost two orders for SD. This means
that the reversal process principally depends on the precession of
the magnetic dipole, and can not be described by a simple two-state
model. This result gives the quantitative substantiation for the
idea to use the tilted magnetic field to speed up the reversal
process \cite{tilt}, and also to use additional weak perpendicular
magnetic field for the same purpose \cite{perp}, which actually
leads to the tilt of the aggregate magnetic field. To roughly
estimate the probability of nonswitching of a dipole by one pulse
with the duration $t_p$, the probability density of switching times
can be considered as Gaussian with the mean $\tau$ and SD $\sigma$.
Then the probability of nonswitching is: $p$=$\frac{1}{2}{\rm
erfc}((t_p-\tau)/\sqrt{2}\sigma))$. For $T$=300 K at the minimum of
$\sigma$ we get $p$=$10^{-48}$ even for $\theta$=$0^\circ$. However,
for $\theta$=$0^\circ$,$5^\circ$ at the minimum of $\tau$ we get
unacceptably large probabilities $0.001$ ($p$=$10^{-56}$ for $\theta$=$45^\circ$).

In Fig. 3 the MRT and SD are presented versus the angle between
anisotropy axis and external magnetic field $\theta$ for different
values of damping $\alpha$, $t_p$=8 ns and $T$=300 K. For
$\theta\to$0, smaller values of damping lead to larger values of
both $\tau$ and $\sigma$ as it must (large values of $\tau$ and $\sigma$ for $\theta\to$0
and $\alpha$=0.01 mean that in this range of parameters the noise-induced reversal occurs).
For larger angles, however, the MRT nearly coincide, while the SD is smaller for smaller values of
$\alpha$. To understand why it is so, let us plot the trajectories
of the magnetization for $\theta$=$0^\circ,5^\circ,45^\circ$, see
the inset of Fig. 3 for $\alpha$=0.1, $t_p$=15.7 ns. One can see
that the precession is largest for $\theta=0^\circ$, for
$\theta$=$5^\circ$ the number of turns is smaller, and for
$\theta$=$45^\circ$ the crossing of the boundary
$\overrightarrow{M}[0,M_y,M_z]$ occurs even without precession. This
explains why the reversal in the latter case happens much faster than
for $\theta$=$0^\circ$ and has little dependence on $\alpha$ in the
limit $\alpha$$\ll$1, see Fig. 3. Since the length of the path is
nearly the same for different $\alpha$ (the MRT nearly coincide) and
the noise intensity is proportional to the damping, this obviously
leads to smaller SD for smaller $\alpha$.

In Fig. 4 the MRT and SD versus angle $\theta$ are presented for
different values of the external magnetic field amplitude. It is
seen that starting from values $20^\circ-30^\circ$ and up to $\sim
70^\circ$ there are flat minima of MRT and SD that corresponds to
the range of angles where the fastest and the most reliable
reversals are realized. For the temperature $T=300{\rm K}$ the
reversal occurs even up to the amplitude value $H_0=3 {\rm kOe}$,
which is two time smaller than the static coercive field $H_c$.

In the present paper the effect of noise on the reversal of a
magnetic dipole has been investigated on the basis of computer
simulation of the Landau-Lifshits equation. It has been demonstrated
that at the reversal by the pulse with sinusoidal shape, there
exists the optimal duration, which minimizes the mean reversal time
(MRT) and the standard deviation (SD, jitter).  Also, both the MRT
and the jitter significantly depend on the angle between the
reversal magnetic field and the anisotropy axis. At the optimal
angle the MRT can be decreased by 7 times for $\alpha$=1 and up to 2
orders of magnitude for $\alpha$=0.01; the jitter can be decreased
from 1 to 3 orders of magnitude (for $\alpha$ from 1 to 0.01) in
comparison with the uniaxial symmetry case. For optimal
angles the SD decreases with decrease of the damping. It has been
demonstrated that fluctuations can not only decrease the reversal
time, as it has been known before for the magnetic systems and is
correct for small angles only, but it can also significantly
increase the reversal time.

\end{document}